\documentclass[12pt]{article}
\input{epsf.tex}
\begin{document}
\begin{center}
{\bf ON PROBLEMS OF EXPERIMENTAL DETERMINATION OF RELIABLE
VALUES OF NUCLEUS PARAMETERS AT LOW EXCITATION ENERGY --
$^{60}$Ni AS AN EXAMPLE
}\\\end{center}
 \begin{center}
{\bf A.M. Sukhovoj, V.A. Khitrov}\\
\end{center}
\begin{center}
 {Joint Institute for Nuclear Research, Dubna, Russia}
\end{center}
\begin{abstract}The reanalysis of the published
experimental data from reaction $^{59}$Co(p,2$\gamma)^{60}
$Ni was performed.
The region of the most probable values of level density
and radiative strength functions of cascade
gamma-transitions was determined.
The obtained data were rather precisely approximated by
the V.M. Strutinsky model and semi-phenomenological model
 -- for strength functions.
The region of appearance and magnitude of maximal
errors of the calculated cross-section of nucleon emission
in evaporation spectra in traditional methods of their
analysis were determined as well.
There was for the first time obtained methodically
correct information on the radiation strength function
of primary gamma-transitions in diapason of neutron
binding energy with averaging over large set of initial levels.
\end{abstract}

\section{Introduction}\hspace*{16pt}

The only goal of an experiment is to get new experimental
information. But, for the further development of science,
especially fundamental one, this information must be more
reliable than that obtained earlier.
Accounting for inevitability of appearance of different
errors and mistakes in this process, the most importance
is their revealing and rejection.
This process requires one to perform new experiments by
the methods not used earlier and to reveal obvious errors
in the performed earlier analysis of experimental data,
for example.
  
This problem appears itself most sharply in low energy nuclear
physics. Development of this branch of fundamental science
is impossible without using different hypotheses and
notions of nuclear properties determined, with the rare
exception, only experimentally. Moreover - only in the
form which requires mathematical treatment of different
extent of complexity, the following physics interpretation
and model description.

To the all enumerated above completely corresponds the
situation with investigation of nucleus properties at
small (up to $\approx 10$ MeV) excitation energy of arbitrary
nucleus. Physically it is set by nucleon binding energy
$B_n$ in a nucleus. Complete and correct understanding of
properties and parameters of a process appeared by this
requires one to determine with good precision the level
density of a nucleus $\rho=D^{-1}$ with given quantum
parameters at any excitation energy and probability (width)
$\Gamma$ of the reaction products emission (strength
function  $k=\Gamma/(D E^3_\gamma A^{2/3})$ for
gamma-quanta).

There are no ways for complete achievement of this goal --
resolution of existing spectrometers does not allow one to
distinguish fixed nuclear levels lying by several MeV
above the ground state and, for example, by several keV
above $B_n$.

Therefore, the values $\rho$ and $k$ can be obtained only
from solution of reverse task of mathematical analysis --
determination of corresponding parameters from the
experimentally measured function $S$ (spectrum or
cross-section). This solution in most cases is badly
stipulated and ambiguous. Id est, the errors
$\delta \rho$, $\delta k$ cannot be equal to zero even at
zero experimental uncertainty of function $\delta S$.
Moreover, the shape of their functional relation is set a
priory on the basis of existing notions.
As a result, the determined parameters of function
$S=f(\rho,\Gamma)$ contain usually more or less not
removable systematical error. 

Intensity of spectra of nuclear reaction products measured
without coincidences with other products (one-step
reactions) is determined by product of $\rho$ and $\Gamma$.
This circumstance increases relative error of
determination of any of these parameters in case if
another parameter was set independently, for example,
by means of some nuclear model. Really, at present this
possibility can be realized only with unknown systematical
error. A lack of further progress in development of the
methods for experimental determination of $\rho$ and
$\Gamma$ in one-step reactions means, in practice, that
the potential of their possibilities is exhausted.

At the same time, real progress in experimental
determination of $\rho$ and $\Gamma$ can be achieved only
from comparison of the data of independent experiments.
This can be made by analysis of results of multi-step
reactions (registration in coincidence of intensity of
cascades from two and more successive nuclear reaction
products). Just study of these cascades terminating at
one or several final levels allowed one to realize
independent model-free method for determination of maximum
reliable values of $\rho$ and $k$.
As it was shown in \cite{{YaF7210}}, only registration of
cascade gamma-quanta in such variant provides maximal
sensitivity of experiment to sought parameters of a
nucleus.

And, accordingly  -- guaranties their lowest systematical
uncertainty with respect to alternative methods for
determination of level density and emission probability
of reaction products.

\section{Dubna method for determination of $\rho$ and
$k$}\hspace*{16pt}

The main and absolutely necessary condition for obtaining
of reliable information from two-step reactions is maximaly
possible test of existing information used in analysis
of experiment. Just practical realization of this
condition allowed one to create the method for simultaneous
model-free determination of $\rho$ and $k$
\cite{Meth1,PEPAN-2005} from experimental spectra by use
of only algorithms, rules and principles of mathematic
and mathematical statistics. Non-observance of these
theses guaranties appearance of principle discrepancies
between the data obtained by different groups from the
one hand and circulation of earlier appeared systematical
errors - from the other hand.

The latest example of this kind is the analysis of the
two-step cascade intensities measured by authors
\cite{ARX1819} in reaction $^{59}$Co$(p,2\gamma)^{60}$Ni.
Experimental spectrum in this as well as in some tens of
analogous cases has the following peculiarities:

(a) it is the sum of two unknown functions corresponding
to near by energy primary and secondary gamma-transitions
in any interval of their values;

(b) the sum of intensities of all the possible two-step
cascades (plus primary transition to the ground state)
always equals 100\%; 

(c) the experimental spectra can be reproduced with the
$\chi^2/f <<1$ value by  infinite number of different
functions $\rho$ and $\Gamma$ from limited interval of
their values;

(d) the relation between change in cascade intensity
$\delta I_{\gamma\gamma}$ and changes in $\delta\rho$ and
$\delta\Gamma$ is non-linear and strongly different for
various energies of cascade gamma-quanta;

(e) the confidence interval for the $\rho$ and $k$ values,
reproducing  $I_{\gamma\gamma}^{exp}$, has practically
acceptable width only by use of the procedure 
\cite{Prim} in analysis of experiment.

Potential possibility to account for enumerated above
specifics of the analysis of the two-step cascade
intensities takes place and in the case \cite{ARX1819}.

1. Level scheme of $^{60}$Ni below excitation energy of
5 MeV includes  \cite{ENSDF} less than 40 secondary
transitions to level $E_f=1.332$ MeV and, correspondingly,
the spectrum  -- narrow peaks. The use of method
\cite{Prim} in this case would allow one to determine
two-step cascade intensities in function on their primary
transition energy with the error caused, in practice, only
by the total error of approximation of peak areas. 

2. Rather essential discrepancies between the experimental
and calculated by authors \cite{ARX1819} spectra of
cascade intensities (for example, for $1.5< E_\gamma<3$)
MeV guarantee, by conditions (b) and (d), their undoubted
difference of unknown amplitude in any other points of the
spectrum.

3. According to \cite{PEPAN-2005}, from combination of
the individual cascade intensities and intensities of
their primary and secondary gamma-transitions, authors of
\cite{ARX1819} could additionally estimate to the first approach
the dependence of radiative strength functions
on energy of levels excited by the primary gamma-transitions. 

The presence of these errors unambiguously requires
reanalysis of the experimental data on the two-step
cascade intensities in $^{60}$Ni. Analogous reanalysis
accounting for the mentioned above specific of the
experiment was earlier performed for $^{57}$Fe, \cite{Fe57},
$^{96}$Mo \cite{Mo96}, $^{172}$Yb \cite{Yb172} and for
some other nuclei. In all these cases real systematical
errors of the values of level densities and radiative
strength functions cited in original articles are
maximally large.  

\section{Conditions of reanalysis}\hspace*{16pt}

Systematical error in determination of $\rho$ and $\Gamma$
from cascade intensities are caused, first of all, by
systematical error of determination of absolute value of
cascade intensity.
According to \cite{TSC-err}, if its value is small enough
($\delta I_{\gamma\gamma}/I_{\gamma\gamma}<50\%$), then
distortions of obtained parameters are also comparatively
small. Their further reduction can be provided by careful
choice of additional experimental data involved in
analysis. 

The total intensity of two-step cascades was renormalized
from the data \cite{ARX1819} and accepted to be equal to
33 events per 100 decays for 
$0.5<E_\gamma<9.5$ MeV. (For example, a portion of the most
intense cascades in near-magic nuclei $^{144}$Nd and
$^{200}$Hg equals accordingly 32\% and 35\%
\cite{Nd144,Hg200}).  By this, intensity of cascades in
interval about $E_\gamma=8.75$ MeV was changed by us to
intensity in interval $E_\gamma=1.25$ MeV because of
presence of undoubted error in \cite{ARX1819}.

The distribution of number of decaying initial cascade
levels was taken in  accordance with \cite{ARX1819} equal
to 39.6; 3.7 and 4.5\% for spins  $J^\pi=3,4^{-}; 1,2^{-};
2,3,4^{+}$ correspondingly (for orbital moment of captured
protons $l=0-2$). In analysis below $E_{ex}=4.02$ MeV
were used known scheme of level density and modes of their
decay. Parameters of the Fermi-gas model -- one of the
variants of initial level density in random iteration
process 
\cite{Meth1,PEPAN-2005} 
for determination of $\rho$ and $\Gamma$ -- were
taken from \cite{PRC76}.
Typical example of the best fit of cascade intensity is
shown in Fig. 1.
 
\begin{figure}\begin{center}
\vspace{3cm}
\leavevmode
\epsfxsize=10cm

\epsfbox{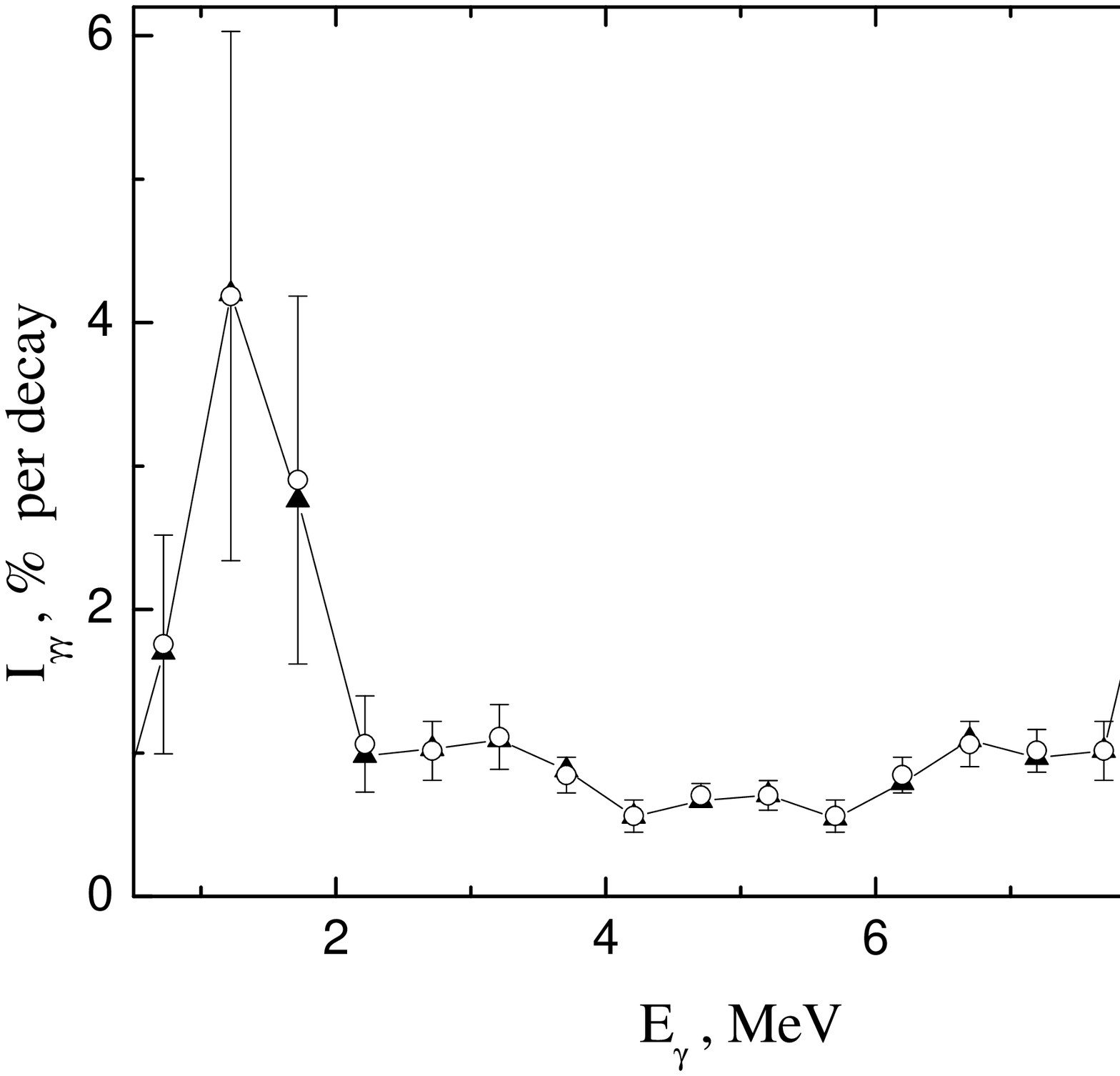} 
\vspace{-3cm}

Fig. 1. Typical approximation of intensities of two-step
cascades to the first excited state of $^{60}$Ni.
Points with errors -- from \cite{ARX1819}, curve with
triangles -- the best fit.
\end{center}
\end{figure}
The total radiative width for initial levels with
enumerated above spins and parity was taken equal to
1.6 eV in accordance with \cite{BNL}. The value of spacing
$D$ between s-resonances for $^{59}$Ni was taken from the
same compilation.
The part of levels of negative parity was additionally
varied in iterative process with respect of initial one
(it linearly decreases from 50\% for $B_n$ to 0 -- at
$E_{ex}=4.02$ MeV at the first step of calculations).
All existing now experimental data on intensities of
two-step cascades in more than 50 nuclei from $^{28}$Al
to $^{200}$Hg demonstrate negligibly small role of pure
quadruple transitions. Therefore, only dipole gamma-quanta
are taken into account in calculation what does not
exclude possibility of presence of mixture of multipoles $M1+E2$ and $E1+M2$.

\begin{figure}\begin{center}
\vspace{3cm}
\leavevmode
\epsfxsize=12cm

\epsfbox{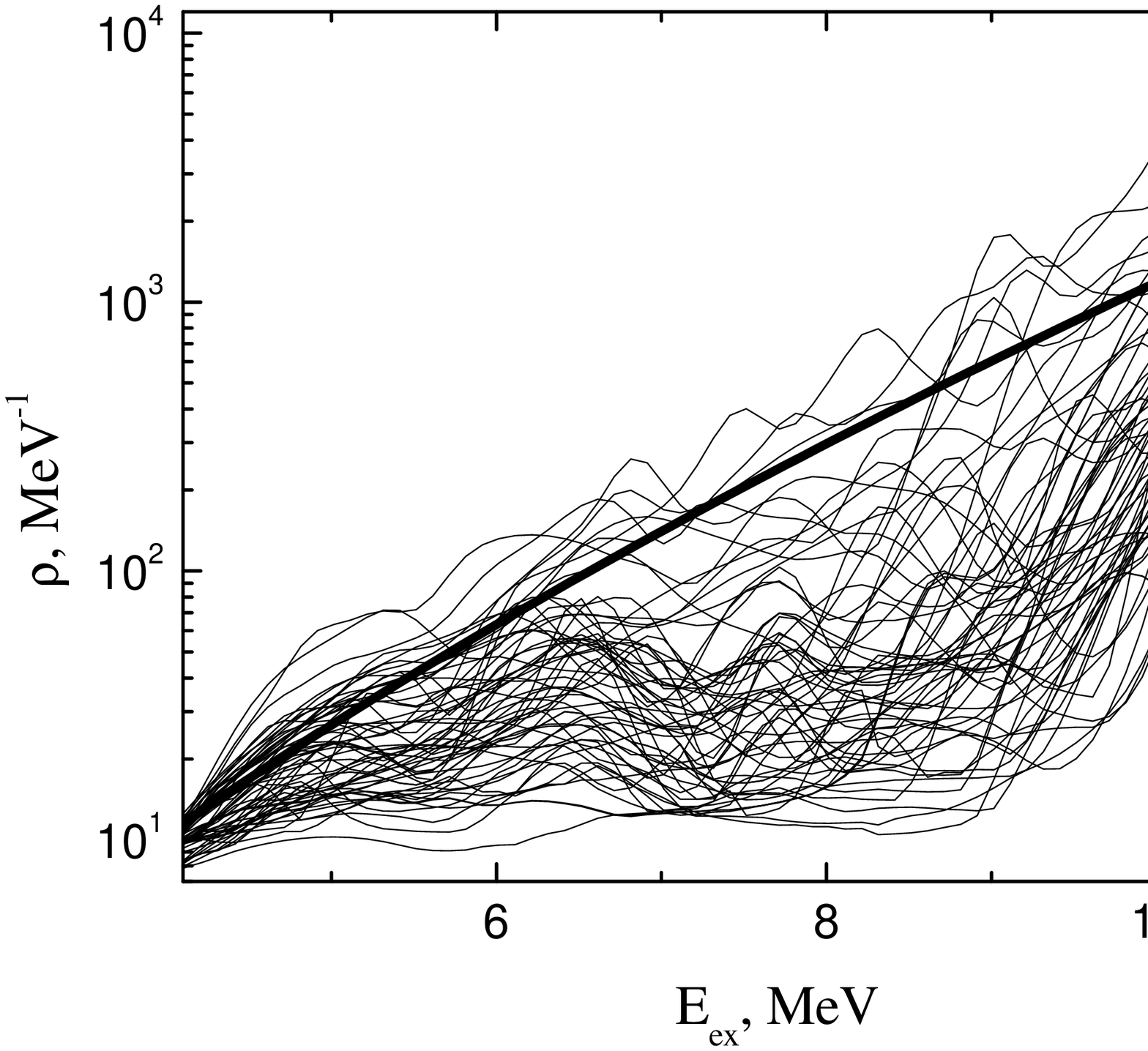} 
\vspace{-4cm}

Fig. 2. Thin curves -- the spectrum of random values of
the level density function ($1 \le J \le5$), reproducing
the data \cite{ARX1819} with $\chi^2/f<0.05$.
Thick curve  -- model \cite{BSFG}.
\end{center}
\end{figure}

Unlike \cite{ARX1819} (determinate by authors variation of
subjectively chosen strength functions), we used the
described in \cite{Meth1,PEPAN-2005} procedure of
multiple distortion of the initial $\rho$ and $\Gamma$
values by small-amplitude random functions with
accumulation of changes decreasing $\chi^2$.
The sets of $\rho$ and $k$ providing
such quality of approximation are given in figures 2 and 3.
\begin{figure}
\begin{center}
\vspace{3cm}
\leavevmode
\epsfxsize=12cm

\epsfbox{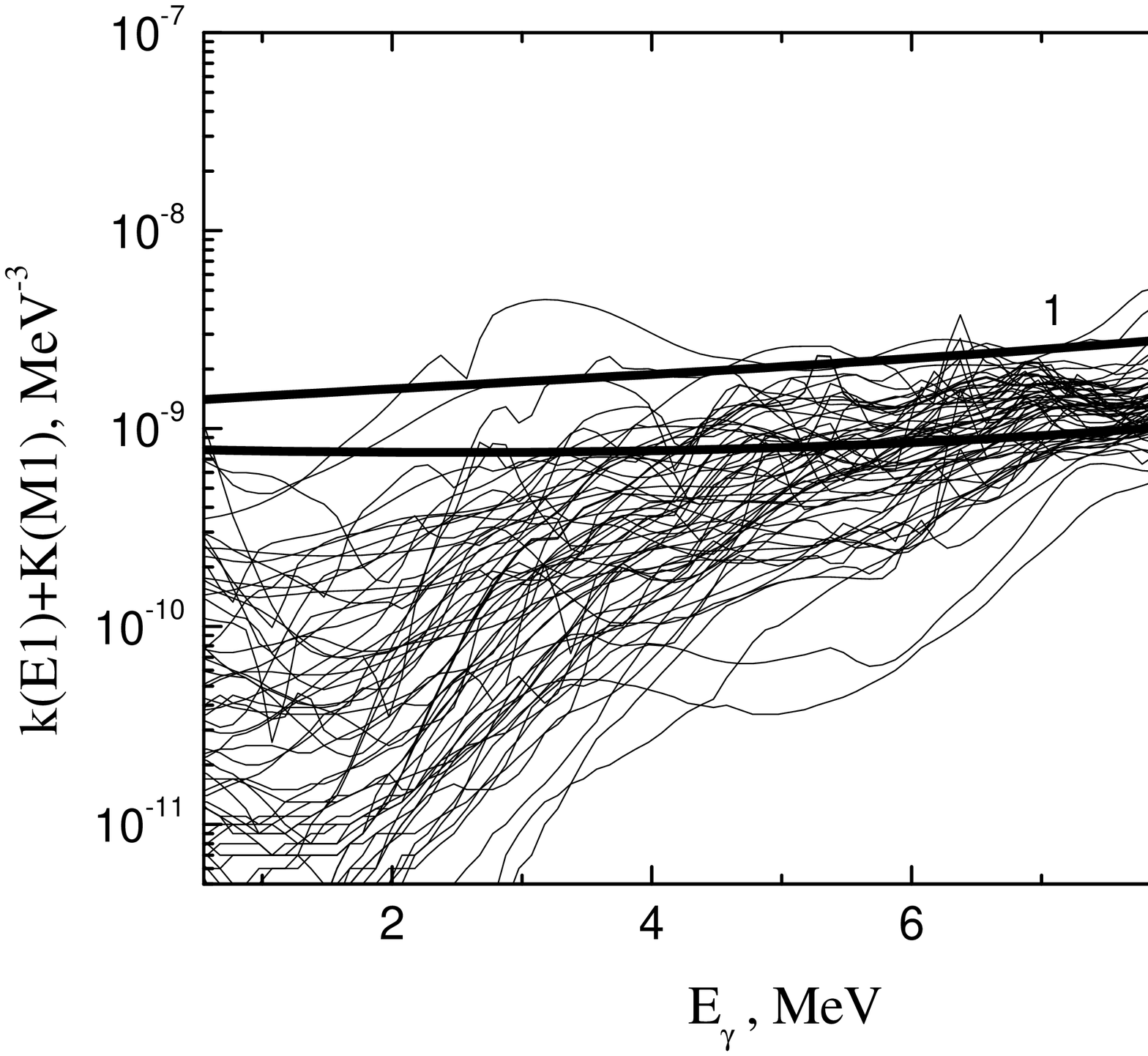} 
\vspace{-4cm}

Fig. 3.  The same, as in Fig. 2, for $k(E1)+k(M1)$.
Curve  1 -- model \cite{Axel},  curve 2 -- \cite{KMF} in
sum with $k(M1)$=const.
\end{center} 
\end{figure}

The dispersion of strength functions and level density for any
cascade quantum and excitation energies is very large
(by one or two orders more than the data \cite{Meth1,PEPAN-2005}).
Its mean-squared value in some energy intervals exceeds the
average $\rho$ and $k$. In these cases, the errors ``down"
in figures 4 and 5 are absent.
Of course, 
each of random functions $\rho$ and $k$ 
 has errors caused by uncertainty
of experimental data
and different reasons for appearance of fluctuations of
widths of primary and
secondary transitions.
But, in difference of principle from \cite{ARX1819},
the best values of
$\rho$ and $\Gamma$ include the dependence of $\rho$ and
$k$ on nuclear
structure at $E_{ex} \le B_n$ to the maximum extent.
Such possibility is completely absent in data treatment \cite{ARX1819} .

The data presented in Fig. 3 show that obligatory
accounting for the obtained in Oslo low energy ``tail"
of strength functions is not required for description of
cascade intensity. But, it must be noted that the errors
of the data presented in figures 2 and 3 are maximal for
the region of excitations lying above  $E_{ex}\sim 9-10$
MeV owing to small statistics and use of the least
informative experimental spectrum  (relatively to possible
cascade intensities in function on their primary
transition energy \cite{Prim}). Due to this reason they
are of small efficiency for determination of the $\rho$
and $k$ values corresponding to this energy. 

Enormous dispersion of the data in figures 2 and 3 is
unambiguously caused by the fact that the calculated by
use of these data cascade intensity with primary and
secondary gamma-transitions with energies from the same
interval really correspond to combinations of analogous
unknown experimental functions $I_{exp}$ and arbitrary
unknown error  $\delta I$: 
\begin{eqnarray}
I^{prim}_{cal}=I^{prim}_{exp}+\delta I\\\nonumber
I^{sec}_{cal}=I^{sec}_{exp}-\delta I.
\end{eqnarray}

As a result, this brings to deviations of $\rho$ and $k$
with different sign and maximal by module for
$I_{exp}=\delta I$.
\begin{figure}\begin{center}
\vspace{2cm}
\leavevmode
\epsfxsize=10cm

\epsfbox{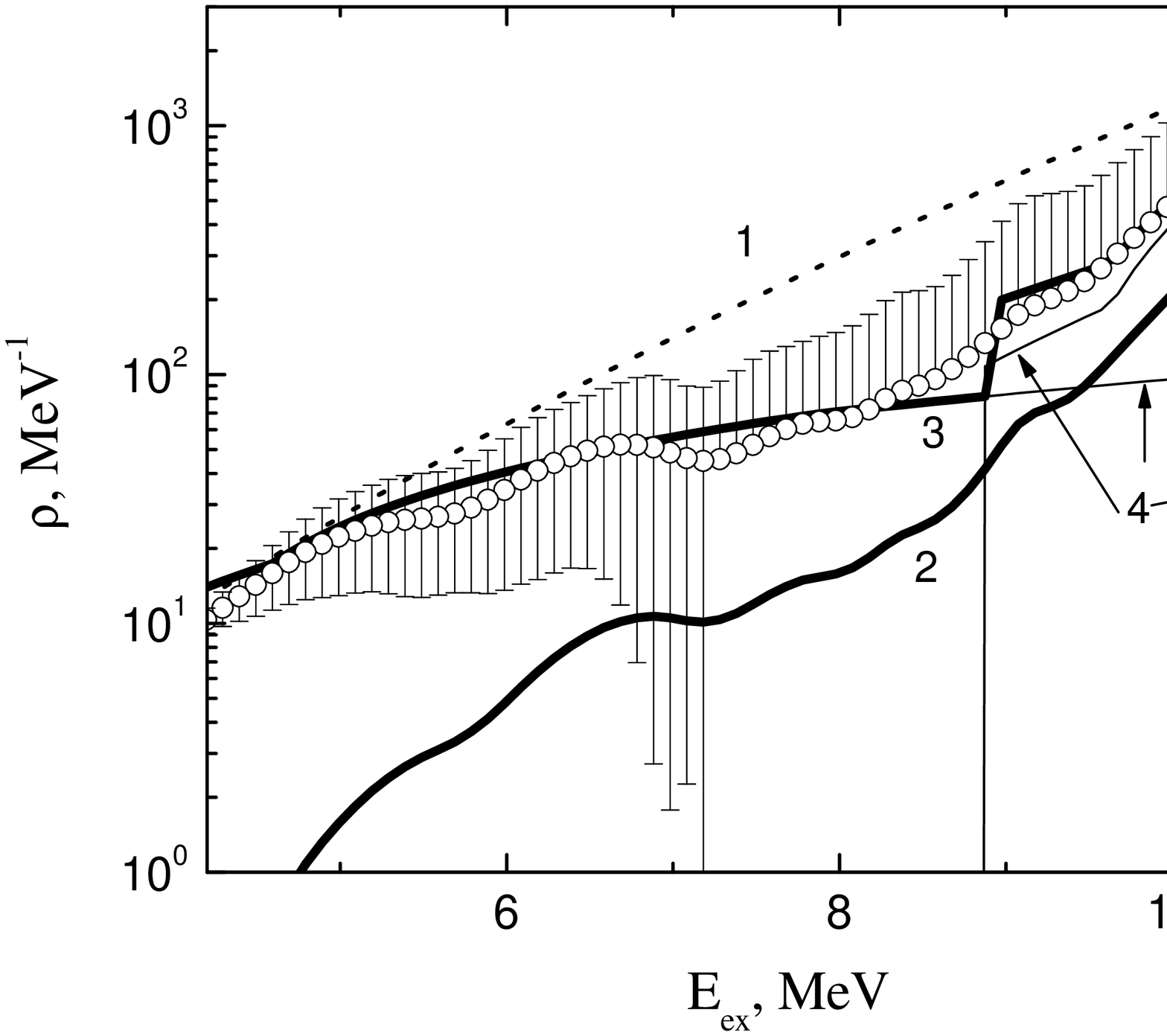} 
\vspace{-3.0cm}

Fig. 4. Approximation of the mean level density from
Fig. 2 by model \cite{Strut}.
Points with errors -- mean value with the mean squared
deviation.  
(The mean squared distribution of the data of Fig. 2 is
given only if it does not exceed the average).
Curve 1 -- model \cite{BSFG},
curve 2 -- fitted value of negative parity level density,
curve 3 -- the best fit,
curve 4 -- partial densities  of 2-, 4- and
6-quasi-partcle excitations.

\end{center}
\end{figure}
Conclusions about parameters $\rho$ and $k$ in this
situation inevitably have essential uncertainty.
Nevertheless, concrete conclusions on gamma-decay process
and its parameters can be made from model approximation of obtained here
$\rho$ and $k$ and for this nucleus.

\section{Model approximation of $\rho$ and $k$}\hspace*{16pt}

Model approximation of functional dependences
$\rho=\phi(E_{ex})$ and $k=\psi(E_\gamma,E_{exp})$ was performed
below for the averages of random functions Figs. 2 and 3
and, therefore, its concrete results are to be considered
as rough enough.
That is why, approximation of $\rho=\phi(E_{ex})$ was
performed in the simplest variant $K_{coll}=$const,
but with accounting for influence of shell inhomogeneities
of one-particle spectrum on parameter  $g$.

In some cases, as a minimum it is possible to determine
a sign of corresponding uncertainty. So, a lack of the
data on extent of dependence of $k=\psi(E_\gamma,E_{ex})$
on excitation energy
rather essentially increases error in determination of
$\rho$ and $\Gamma$. However, it follows from comparison
of variants for their determination  \cite{Meth1} and
\cite{PEPAN-2005} that the obtained for nickel and
presented in figures 4 and 5 $\rho$ values are
overestimated  and  $k$ -- underestimated.

Approximation of $k=\psi(E_\gamma)$ was performed within
frameworks of semi-phenomenological model \cite{appr-k}.
The results are shown in figures 4 and 5.

\begin{figure}[h]
\begin{center}
\vspace{3cm}
\leavevmode
\epsfxsize=10cm

\epsfbox{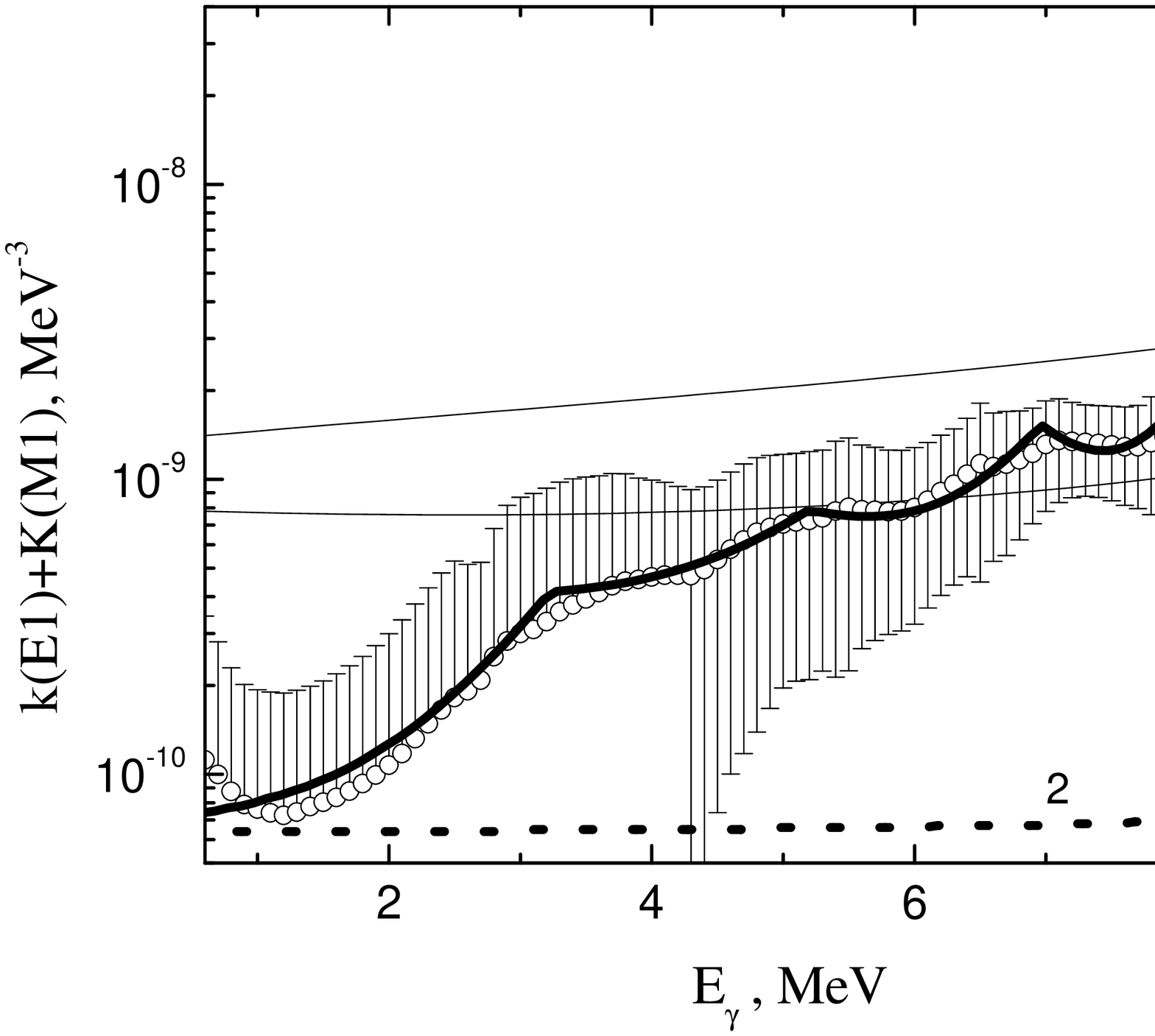} 
\vspace{-3cm}

Fig. 5. The same, as in Fig. 4, for $k(E1)+k(M1)$.
Curve 1 -- the best fit, curve 2 --
contribution of model \cite{KMF}.
\end{center}
\end{figure}

As a result, one can accept that the analysis described
above allowed one to determine main properties of nucleus
$^{60}$Ni. The discrepancies of the $\rho$ and $k$ values
with the Dubna data for this nuclei are to
high extent caused by systematical errors of used here ``alternative"
methods of determination of the same parameters.  

{\sl Table~1.\\ } Parameters of approximation of level
density and radiative strength functions for different
nuclei: the coefficient of change of square of nuclear
temperature $\kappa $ and contribution $w$ of model
\cite{KMF} in the summed  strength function. $E_1$ --
position of local peak and its amplitude $P_1$
 (multiplied by $10^{-7}$), $\alpha$ - velocity of
 decrease of amplitude as primary transition energy
 decreases. Parameter of level density  $a$, coefficient
 of collective enhancement of level density
 $K_{\rm coll}$ and thresholds $U$ of break of the second
 and third Cooper pairs obtained in variant 
\cite{PEPAN-2006} accounting for shell inhomogeneities
of one-particle spectrum.

\begin{center}
\begin{tabular}{|l|l|l|l|l|l|l|}  \hline
 Parameter           & $^{60}$Ni &$^{74}$Ge & $^{96}$Mo &  $^{114}$Cd & $^{118}$Sn & $^{124}$Te\\\hline
$\kappa $            & 0.0002 &  0.14(6) & 0.41(16)  &  0.18(9)    & 0.04(25)  & 0.18(3)\\    
$w $                 & 0.011 &  0.25(4) & 0.16(3)   &  0.10(4)    & 0.01(4)   & 0.56(6)\\
$E_1$, MeV           & 5.10(2) &  5.4(1)  & 5.8(2)    &  5.7(1)     & 5.0(1)    & 7.3(1)\\
$P_1$                & 0.37(2) &  5.3(2)  & 4.0(6)    &  9.6(36)     & 8.2(6)    & 7.2(7)\\
$\alpha,$ MeV$^{-1}$ & 1.03(2) &  0.59(35)& 0.74(21)   &  0.89(6)    & 0.90(8)   & 0.80(9)\\
$a$, MeV$^{-1}$      & 6.16 &   9.96   & 11.       &  13.        & 13.3      & 15.6\\
$K_{coll}$           & 12 &   17    &  6.7       &  13         & 4.5       & 15\\
$U_2$, MeV           & 8.9 &   5.9    & 3.7       &  3.8        & 4.7       & 2.8\\
$U_3$, MeV           & 10. &   8.8    & 6.5      &  5.9        & 4.3       & 6.6\\\hline
\end{tabular}\end{center}


Unlike the data on $k$ derived from two-step cascade
intensities, reproduction of form of energy dependence
in case under consideration by model \cite{appr-k}
requires one to take into account not less than four local
peaks (or corresponding continuous distribution,
exponentially decreasing when primary gamma-transition energy
decreases). Large values of $K_{coll}$ and break threshold
of the second Cooper pair  $U_2$ allow one to relate the
observed dependence and its parameters with
gamma-transitions between levels with large phonon
components of wave functions. The main difference of the
data for  $^{60}$Ni from the data  \cite{appr-k} consists
in significant averaging  \cite{ARX1819} of strength
functions over large set of initial cascade levels.
The distribution shown in Fig. 5 can be combined from
some peaks with exponentially decreasing tails.
This result can be interpreted as enhancement of
gamma-transitions at decay of fixed resonance with
primary excitation of some intermediate level groups in different
energy intervals. Test of this hypothesis needs in both
additional experiments, analogous to \cite{ARX1819},
and correct from the point of view of principles of
mathematics and mathematical statistics treatment of their
results. I. e., adequate by these conditions to
\cite{PEPAN-2005}.

\section{Sources of systematical errors of different
methods}\hspace*{16pt}

1. Large volume of the experimental $\rho$ and $k$ values
was obtained from analysis of the total gamma-spectra and
presented in publications by Norwegian collaboration.
There are two different of principle differences between
their data and Dubna data:

(a) absolute absence of a something essential deviation of
$\rho$ from ``smooth" dependence and 

(b) absolute absence of some estimations of systematical
error of the $\rho$ and $f=k A^{2/3}$ values presented by
authors  \cite{NIM}. In particular, the authors did not
show a precision of determination of systematical
uncertainty of absolute intensity of the total
gamma-spectra measured by them for different excitation
energies. 

But, as it was shown in \cite{TotSpe}, obtaining of the
reliable $\rho$ and $f$ values requires lesser
(most probably, much lesser) their total uncertainty
(for example, 1\%)  for any gamma-quanta energy.
Without such proof, the observed discrepancy of $\rho$
and $k$ between \cite{PEPAN-2005} and \cite{NIM} can be
relate, first of all, with the systematical errors
of using the method \cite{NIM} for experimental total gamma-spectra
data treatment.

2. Level density in $^{60}$Ni, obtained from the spectra
of evaporation nucleons in  \cite{PRC76}, very strongly
differs from function $\rho$ precisely reproducing 
cascade intensities (Fig. 2). Most probably, the source of
error is the error of calculation of differential cross
section of product emission
$\frac{d\sigma}{d\varepsilon}$ in reaction with non-zero residual
excitation energy of final nucleus.

Up to now it is calculated only by optical model of a
nucleus and only in the frameworks of unvereficated
hypothesis on independence of partial width $\Gamma$ of
emission of the nuclear reaction product on nuclear
structure. Id est, the known parameters of interaction
of reaction product with non-excited nucleus are
extrapolated to the region where occurs sharp change of
its properties \cite{PRC76,Vona83}.

In this region, according to
quasi-particle-phonon model of nucleus, nucleon does not
interact with nucleus being in the state of quasi-particle vacuum excitations,
but interacts with nucleus in the state, for example,
``quasi-particles $\otimes$ quadrupole phonon" or more complicated structure.
Moreover -- with different fragmentations at different
nuclear excitation energy \cite{MalSol}.

Conclusion about dependence of $\Gamma$ on nuclear
structure (id est, excitation energy of final nucleus)
unambiguously follows, for example, from \cite{PEPAN-1972}.
And it fully agrees with the obvious axiom that any
extrapolation of theoretical model in unstudied region
of excitations has an error which can be determined only
experimentally.

According to Hauser-Feshbach notion, the cross-section
under consideration is determined by sum over initial and
final levels of products of type 
$\Gamma_b(U,J,\pi,E,I,\pi)\rho_b(E,I,\pi)$  for the final
reaction product $b$ \cite{PRC76,Vona83}:

\begin{eqnarray}
{\frac{d\sigma}{d\varepsilon_b}(\varepsilon_a,\varepsilon_b)=}
\sum_{J\pi}\sigma^{\mathrm{CN}}(\varepsilon_a)\,\frac{\sum_{I\pi}
\Gamma_b(U,J,\pi,E,I,\pi)\rho_b(E,I,\pi)}{\Gamma(U,J,\pi)}
\end{eqnarray}
where
\begin{eqnarray}
\lefteqn{\Gamma(U,J,\pi)=\sum_{b^\prime}\left(\sum_k \Gamma_{b^\prime}(U,J,\pi,E_k,I_k,\pi_k)\,+\right.}\\
&&\left.\sum_{I^\prime\pi^\prime}\int_{E_c}^{U-B_{b^\prime}}dE^\prime\,
\Gamma_{b^\prime}(U,J,\pi,E^\prime,I^\prime,\pi^\prime)\,
\rho_{b^\prime}(E^\prime,I^\prime,\pi^\prime)\right).\nonumber
\end{eqnarray}

It follows from (2) and (3) that at presence of unknown
error $\delta$ of the calculated $\Gamma$ width value,
experimental cross-section $\frac{d\sigma}{d\varepsilon}$
can be precisely reproduced only by using level density
with adequate systematical error.

If the calculated partial $\Gamma_{cal}$ and experimental
$\Gamma_{exp}$ widths obey to 
correlation $\Gamma_{cal}=\Gamma_{exp}(1+\delta)$,
then the level density $\rho_{cal}=\rho_{exp}/(1+\delta)$
must be used in calculation for reproduction of
cross-section (2). Of course, unknown relative error $\delta$
of calculated width depends on excitation energy of final
nucleus and can depend on spin of levels which are
connected each to other at $b$ reaction product emission.
In this case, the error $\delta$ is the weight average.
If one determines it using relation: 
$(1+\delta)=\rho_{es}/\rho_{2\gamma}$, which connects
level density $\rho_{es}$  determined from evaporation
spectrum \cite{PRC76,Vona83} with density
$\rho_{2\gamma}$, derived from cascade intensity
(fig. 4), then the error of the calculated
cross-section $\frac{d\sigma}{d\varepsilon}$ can be
directly estimated at any excitation energy of final
nucleus. The result is presented in Fig. 6.
The main error of the presented in this figure data is
related to the fact that the $\rho$ value was determined 
directly from two-step experimental spectrum, but not from cascade
intensities for given energy of their primary transition
and because of lack of information on function
$k(E_\gamma , E_{ex})$. However, there are no principle
differences with analogous data for $^{181}$W
\cite{Isinn17b}.
\begin{figure}\begin{center}
\vspace{3cm}
\leavevmode
\epsfxsize=10cm

\epsfbox{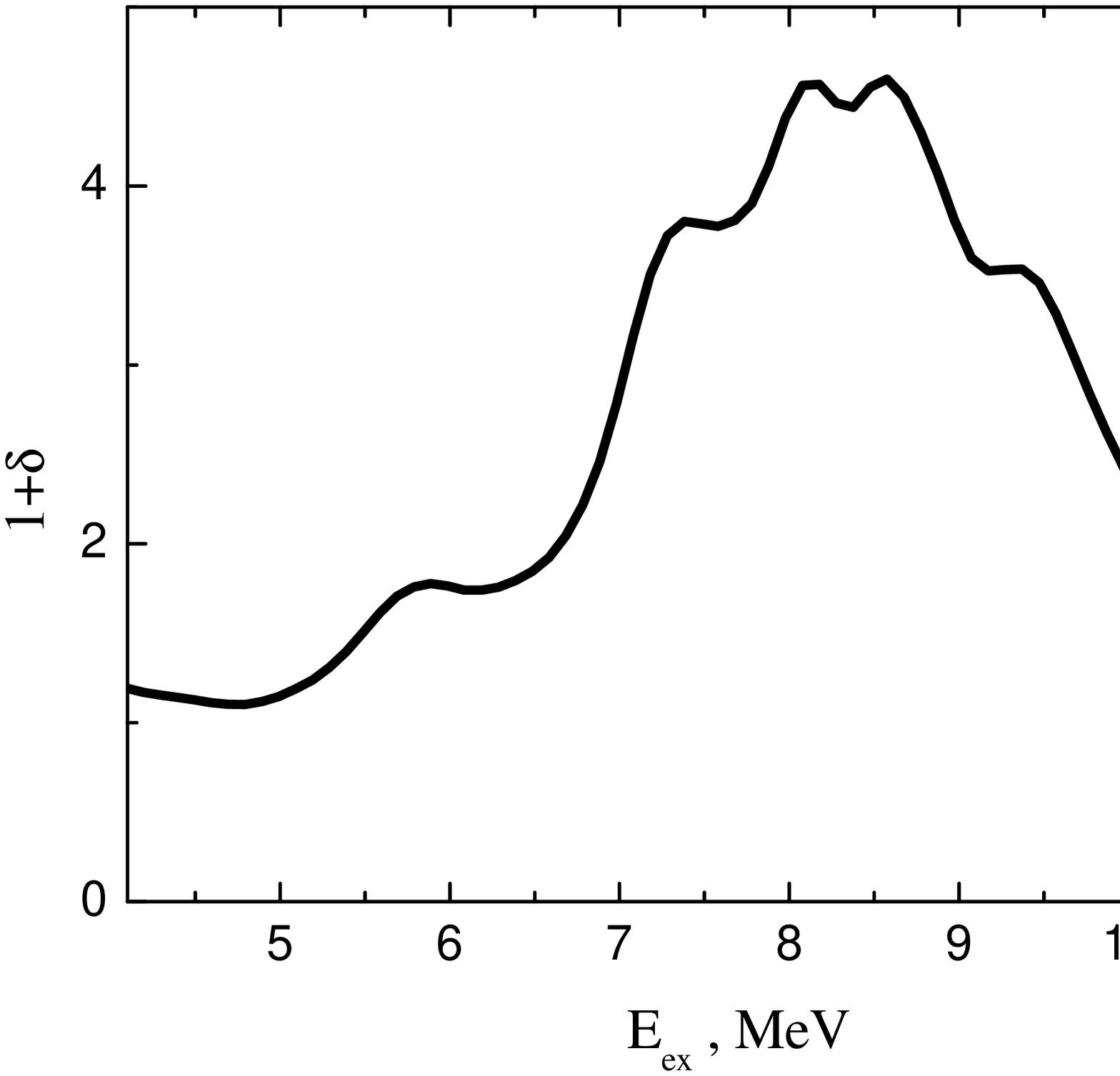} 
\vspace{-2.5cm}

Fig. 6. The degree of enhancement of differential
cross-section of nucleon emission in $^{60}$Ni for
different excitation energy of final nucleus.
\end{center}
\end{figure}

3. The any ordinary errors of the expected amplitude cannot
even in principle distort the form of energy dependence 
of $\rho$ and $k$ determined in correspondence with 
\cite{Meth1,PEPAN-2005}. Serious systematical error in
this case can have only more fundamental character.
For instance, connected with lack of the model of
gamma-decay of highly excited levels accounting in
available (for analysis of experimental data) form for
coexistence and interaction of excitations of fermion
and boson type and so on.

\section{Conclusion}\hspace*{16pt}

1. Confirmation of the data on level density derived from
the spectra of evaporation nucleons by analogous data
from the two-step cascade intensities is achieved only
subjectively at serious violations of statements of
mathematics and  mathematical statistics at analysis of
$I_{\gamma\gamma}$.

2. As in the other spherical even-even nuclei, taking
into consideration, break of three Cooper pairs of
nucleons is quite enough for model description \cite{Strut} of level
density in $^{60}$Ni below $B_n$. Collective effects
increase level density by order of magnitude.

3. High degree of averaging of cascade intensities over
their initial levels allows one to expect considerable
averaging of strength functions.
In particular, the strength function of the primary
gamma-transitions to the first two-phonon state
$E_f=2.505$ MeV of this nucleus is increased,
as a minimum, by order of magnitude with respect to the
model \cite{Axel,KMF} predicted values. It is possible also that the
increased gamma-transitions following decay of different
initial cascade levels selectively excite groups of
levels of collective structure at different excitation
energy. Id est, the parameters of model approximation of
the strength function determined from ``averaged resonances"
can strongly differ from analogous strength functions of
decay of the only initial cascade level.

4. Precise description of the two-step gamma-cascade
intensities does not require increase of strength
functions at decrease of gamma-transitions energy.
This effect is easily quantitatively
explained even by small errors \cite{Mo96norm} at
normalization of the total gamma-spectra at different
excitation energy of spherical nucleus.

5. The notion of authors, for instance, \cite{PRC76} that
the approximate equality of level densities derived by
them from the spectra of evaporation nucleons in different
nuclear reactions is a proof of absence of systematical
errors is obviously mistaken. In practice, these very big
errors of the calculated value of cross-section are
completely correlated.  Id est, they are caused by
strong influence of structure of excited levels of final
nucleus but not -- by any errors of the used optical
potentials. 

6. Because dominant proton capture occurs \cite{ARX1819} to the level
$\pi=-$, then this strongest increase of strength function is first
observed for the primary $E1$-transitions of the compound-state decay
to final two-phonon level. 



\begin{thebibliography}{99}
\bibitem{YaF7210}
 A.M. Sukhovoj, V.A. Khitrov, W.I. Furman,
Phys. At. Nucl.  {\bf 72(10)}, (2009) 1759.
\bibitem{Meth1}
E.V. Vasilieva, A.M. Sukhovoj, V.A. Khitrov, Phys.  At. Nucl. 
{\bf 64(2)} (2001) 153, (nucl-ex/0110017).
\bibitem{PEPAN-2005}
	A.M.  Sukhovoj, V.A.  Khitrov, Phys. Particl.
	and Nuclei, {\bf 36(4)} (2005) 359.
\bibitem{ARX1819}
	A. Voinov et al.,  Phys.Rev. C {\bf 81} (2010) 024319.
\bibitem{Prim}
S.~T.~Boneva, V.~A.~Khitrov, and A.~M.~Sukhovoj,
Nucl.  Phys.  A
{\bf 589}, (1995) 293.
\bibitem{ENSDF} http://www.nndc.bnl.gov/nndc/ensdf; http://www-nds.iaea.org.
\bibitem{Fe57} A.M. Sukhovoj, V.A. Khitrov, Li Chol, Pham Dinh Khang, Vuong Huu Tan,\\ 
Nguyen Xuan Hai,
In {\it  Proceedings of the XIII International Seminar on Interaction
of Neutrons with Nuclei,  Dubna,  May 2006},
E3-2006-7, Dubna, 2006, p. 72, nucl-ex/0508007.
\bibitem{Mo96}
 A.M. Sukhovoj, V.A. Khitrov,
Phys. At. Nucl.  {\bf 72(9)}, (2009) 1426.
\bibitem{Yb172}  V.A. Khitrov, A.M. Sukhovoj, Pham Dinh Khang, Vuong Huu Tan, Nguyen Xuan Hai,
in  {\it Proceedings of the XI International Seminar on Interaction
of Neutrons with Nuclei,  Dubna, 22-25 May 2003},
E3-2004-9, Dubna, 2004, p. 107, nucl-ex/0305006.
\bibitem{TSC-err}
	V.~A.~Khitrov,  Li Chol, and A.~M.~Sukhovoj, in
	{\it Proceedings of the XI International Seminar on Interactions
	of Neutrons with Nuclei,  Dubna, May 2003}, Preprint No.
	E3-2004-9, JINR (Dubna, 2004), p. 98;  nucl-ex/0404028.
\bibitem{Nd144} Yu.P. Popov et al., Izv. Akad. Nauk SSSR, Ser. Fiz.,
{\bf 48(5)}, (1984) 1830.
\bibitem{Hg200}  E.V. Vasilieva et al.,  
	Bull. Rus. Acad. Sci. Phys. {\bf 60}, (1996) 1706.
\bibitem{PRC76} A.V. Voinov et al., Phys.Rev. C {\bf 76} (2007) 044602.
\bibitem{BNL}	S.~F.~Mughabghab, {\it Neutron Cross Sections
	BNL-325}.  V.  1.  Parts A, edited by S.~F.~Mughabhab, M.~Divideenam,
	N.~E.~Holden, (N.Y., Academic Press, 1984).
\bibitem{BSFG}  W. Dilg, W. Schantl,  H. Vonach,  M. Uhl,
	Nucl. Phys. A {\bf 217} (1973) 269.
\bibitem{Axel}  P. Axel,  Phys. Rev. 1962. {\bf 126(2)} (1962) 671. 
\bibitem{KMF}  S.G. Kadmenskij, V.P. Markushev, V.I. Furman,
	Sov. J. Nucl. Phys. {\bf 37}, (1983) 165.
\bibitem{appr-k} 
	A.M. Sukhovoj, W.I. Furman, V.A. Khitrov,
	Physics of atomic nuclei,  {\bf 71(6)}  (2008) 982.
\bibitem{Strut}
	V.M. Strutinsky, in {\it Proceedings of the International
	Congress on Nuclear Physics}
	(Paris, 1958), p. 617.
\bibitem{PEPAN-2006} 
  	A.M. Sukhovoj, V.A. Khitrov,
	Physics of Paricl. and Nuclei, {\bf 37(6)} (2006) 899.
\bibitem{NIM}
	A. Schiller {\it et al}., Nucl. Instrum. 
	Methods A, {\bf 447}, (2000) 498.
\bibitem{TotSpe} A.M. Sukhovoj, V.A. Khitrov,
In {\it  Proceedings of the XVII International Seminar on Interaction
of Neutrons with Nuclei,  Dubna,  May 2009},
E3-2009-36, Dubna, 2010, p. 268, nucl-ex/1009.4761.
\bibitem{Vona83}
	H. Vonach, in
	{\em Proc. IAEA Advisory Group Meeting on Basic and Applied Problems
	of  Nuclear Level Densities\/} (New York, 1983), INDC(USA)-092/L,
	(1983), p.247.
\bibitem{MalSol} 
	 L.A. Malov, V.G. Solov'ev,  Yad. Phys., {\bf 26(4)} (1977) 729.
\bibitem{PEPAN-1972}
 	 V.G. Soloviev, Sov. Phys. Part. Nuc. {\bf 3} (1972) 390.
\bibitem{Isinn17b}
A.M. Sukhovoj, V.A. Khitrov,
Physics of atomic nucleus,  {\bf 73(10)} (2010) 1635.
\bibitem{Mo96norm} A.M. Sukhovoj, V.A. Khitrov,
In {\it  Proceedings of the XVI International Seminar on Interaction
of Neutrons with Nuclei,  Dubna,  May 2008},
E3-2009-33, Dubna, 2009, pp. 203, nucl-ex/0906.3060.
\end{thebibliography}
\end{document}